\documentclass[conference]{IEEEtran}
\usepackage{blindtext, graphicx}
\usepackage{fancyhdr}
\ifCLASSINFOpdf
\else
\fi

\newcommand{\gi}[1]{{#1}^{(i)}}
\usepackage{amssymb}

\usepackage{listings}
\usepackage{xcolor}
\usepackage{graphics}
\usepackage{amsmath}

\colorlet{punct}{red!60!black}
\definecolor{background}{HTML}{EEEEEE}
\definecolor{delim}{RGB}{20,105,176}
\colorlet{numb}{magenta!60!black}

\lstdefinelanguage{json}{
    basicstyle=\normalfont\ttfamily,
    numbers=left,
    numberstyle=\scriptsize,
    stepnumber=1,
    numbersep=8pt,
    showstringspaces=false,
    breaklines=true,
    frame=lines,
    backgroundcolor=\color{background},
    literate=
     *{0}{{{\color{numb}0}}}{1}
      {1}{{{\color{numb}1}}}{1}
      {2}{{{\color{numb}2}}}{1}
      {3}{{{\color{numb}3}}}{1}
      {4}{{{\color{numb}4}}}{1}
      {5}{{{\color{numb}5}}}{1}
      {6}{{{\color{numb}6}}}{1}
      {7}{{{\color{numb}7}}}{1}
      {8}{{{\color{numb}8}}}{1}
      {9}{{{\color{numb}9}}}{1}
      {:}{{{\color{punct}{:}}}}{1}
      {,}{{{\color{punct}{,}}}}{1}
      {\{}{{{\color{delim}{\{}}}}{1}
      {\}}{{{\color{delim}{\}}}}}{1}
      {[}{{{\color{delim}{[}}}}{1}
      {]}{{{\color{delim}{]}}}}{1},
}

\begin{document}
\title{Grapevine: A Wine Prediction Algorithm Using Multi-dimensional Clustering Methods}

\author{\IEEEauthorblockN{Richard Diehl Martinez}
\IEEEauthorblockA{Computer Science \\ 
Stanford University \\
Email: rdm@stanford.edu}
\and
\IEEEauthorblockN{Geoffrey Angus}
\IEEEauthorblockA{Computer Science \\ 
Stanford University \\
Email: gangus@stanford.edu}
\and
\IEEEauthorblockN{Roozbeh Mahdavian}
\IEEEauthorblockA{Computer Science \\ 
Stanford University \\
Email: rooz@stanford.edu}}


%Telephone: (800) 555--1212, Fax: (888) 555--1212}


\maketitle

\begin{abstract}
We present a method for a wine recommendation system that employs multidimensional clustering and unsupervised learning methods. Our algorithm first performs clustering on a large corpus of wine reviews. It then uses the resulting wine clusters as an approximation of the most common flavor palates, recommending a user a wine by optimizing over a price-quality ratio within  clusters that they demonstrated a preference for. 
\end{abstract}


% \fi
\IEEEpeerreviewmaketitle

\section{Introduction}
Wine has incredible diversity; there exist over 10,000 different varieties of wine grapes worldwide, and each can be processed in a hundred thousand unique ways. Sommeliers---those who dedicate their lives to the art of wine tasting---work to craft flavor profiles for the wines they are given to analyze, using their extensive experience to provide nuanced evaluations of countless bottles of wine every year. But the majority of people have neither the time nor the money to try a variety of wines and develop their palate. Typically, the only claim one can make about a given glass of wine is whether or not it was enjoyable, and without the ability to identify one’s taste preferences in wine, it is incredibly difficult for one to discover new wine, and nearly impossible to find wine that directly matches their individual flavor profile.

We hope to develop an algorithm to address both of these issues, becoming a personal sommelier for the user. Our algorithm takes a history of the wine a user has tasted as input, and returns a set of optimal wines for the user to try next, as well as a description of the flavor profile that inspired the recommendations. Thus, the algorithm could become an avenue for the user to confidently explore wine, and understand more quickly what they do and do not like in wine. 

Formally, we define our problem as an unsupervised learning problem. Let $X \in \mathbb{R}^m \times \mathbb{R}^n$ be the design matrix, where $m$ is the number of wines in our dataset and $n$ is the number of features collected for each wine. Additionally, let some $H$ be some vector describing a user's history of wine consumption, where $\gi{h}$ is some wine the user either liked or disliked. Our objective is to cluster each $x \in X$ to $k$ clusters such that $x$ resides in a cluster of wines with similar flavor profile. This clustering is achieved through the use of the k-means and EM algorithms. Then, given a user's history $H$, we seek to provide some high-quality, affordable wine recommendation $w$ that is similar to the wines found in $H$.

\section{Related Literature}
Categorizing and predicting consumer preferences is a difficult task. Given the especially fickle nature of human taste, the effective application of machine learning in recommendation systems has long been studied by researchers and online retailers alike[1]. 
Historically, natural language processing and supervised learning methods have been primarily used to model consumer preference. Support Vector Machines (SVMs) in particular, were long viewed as the gold-standard for predicting the degree to which a product matched with a consumer’s preferences[2]. In order to ascertain the qualities of an item or service, natural language processing and classification methods are often used to extract the relevant information from a corpus of information about a good. In their 2012 paper, Sakai and Hirokawa demonstrate how to extract the main ‘feature’ words out of an article [3]. The method the researchers outline uses six-fold cross validation on a SVM trained on a corpus of documents that have been normalized by term frequency-inverse document frequency (TF-IDF). These methods, which showed 90\% accuracy on test data, have since been employed in algorithms designed to extract the meanings out of subjective product reviews, such as wine reviews [4]. \\ Building on this work, McAuley et. al have similarly shown how using supervised learning methods, features can be extracted efficiently from corpora of texts consisting of 5 million data points, with multiple dimensions along which to measure ‘quality’ of a product [5]. 
In the domain of wine recommendation systems in particular, scholars have relied on supervised methods such as basic least-squares regression modeling. Frank and Kowalski propose employing simple regression to estimate the quality of a wine from the wine’s objective chemical measurements [6]. Using this model, the researchers predicted subjective sensory evaluation from a wine’s chemical composition. Determining the accuracy of these models, however, remains a difficult task. 
Most recently unsupervised methods have begun replacing standards models for recommendation tasks. This shift in paradigm has come from the realization that clustering products into distinct groups makes it possible to increase the accuracy of recommendations on an individual basis. That is, by first modeling the general differences between groups of similar groups, prediction algorithms can then more accurately derive heuristics for the type of product that fits into a user's preferences [7]. This methodology has been applied by companies like Netflix, which recommend movies to users by first looking at similarities between videos, and then selecting an optimal, personalized choice out of this group based on user history [8]. Our algorithm is based directly off of this framework, and is described in the subsequent section.

%\begin{figure}[!t]
%\centerline{\subfloat[Case I]\includegraphics[width=2.5in]{subfigcase1}%
%\hfil
%\subfloat[Case II]{\includegraphics[width=2.5in]{subfigcase2}%
%\label{fig_second_case}}}
%\caption{Simulation results}
%\label{fig_sim}
%\end{figure*}

\section{Dataset and Features}

Our dataset comprises of wine reviews scraped from WineSpectator.com [9]. Each winery had to be scraped preliminarily as well, for each review was only available through querying its winery page. We designed a fault-tolerant system on top of the \texttt{scrapy} library capable of scraping the wineries and their respective reviews. Over the course of several days, the system compiled a list of over 21 thousand wineries and 350 thousand wine reviews from the website. Each raw review object consists of the following properties: metadata, such as the wine’s name, vintage, winery, region, and country;
score, as given by the sommeliers of WineSpectator; the market price; and finally, the review itself.
\begin{figure}[h]
\begin{lstlisting}[language=json,firstnumber=1]
{
    "name": "Chambolle-Musigny Les Cras",
    "url": ...,
    "country": "France",
    "review": "Candied cherry, cinnamon, violet and black currant notes ride the nervy acidity in this crisp red. Turns pinched in the end. Best from 2012 through 2016. 1,500 cases made.",
    "price:": "\$65",
    "score": "84",
    "winery": "Antonin Guyon",
    "vintage": "2008",
    "region": "Burgundy"
}
\end{lstlisting}
\caption{An example JSON object collected from WineSpectator.com}
\end{figure}

We first filtered wines with scores under 80 points; this number is a common benchmark used to determine quality [10], and our ultimate goal is to recommend quality wines. After doing this, we focused our attention on the properties of the 4th property: the review text itself.

Our clustering algorithm is based on the features of each review text. Thus, our feature extractor was a program implemented to process the review strings of each example. To maximize the salience of words in the review text, we preprocessed away punctuation, capitalization, and generic stopwords. 

The feature extractor built the design matrix $X$ to have $m$ rows and $n$ columns, where $m$ is the number of examples in the dataset and $n$ is the number of words in the vocabulary used throughout the entire dataset. Each example $\gi{x}$ is an $n$-vector where each $\gi{x}_j$ is the TF-IDF value of $j^{th}$ word in the vocabulary. As we are operating with the underlying assumption that the words are in sommelier reviews are incredibly precise, we utilize TF-IDF because of its ability to capture the uniqueness of words in the vocabulary, a property essential to the efficacy of our clustering algorithm.

In order to further distill our dataset, we additionally ventured to remove domain-specific stopwords from the dataset. In order to do this, we ran several iterations of the clustering algorithm and collected the indices of the top-25 highest valued elements in the centroids of each cluster. We then mapped these indices back to the vocabulary. Words common across the clusters were collected, and after manual verification, removed if deemed overly generic.
\begin{figure}[h]
\centering
\begin{tabular}{
|p{3cm}||p{3cm}|p{3cm}|p{3cm}|  }
 \hline
 \multicolumn{2}{|c|}{Domain Stopwords} \\
 \hline
 \hline
TANNINS	&	FLAVORS\\\
FLAVOR	&	DRINK\\
WINE	&	FINISH\\
HINTS	&	FRUIT\\
NOTES	&	OFFERS\\
AROMAS	&	STYLE\\
CHARACTER&	HINT\\
BIT		&DRINKABLE\\
PALATE	&	IMPORTED\\
 \hline
\end{tabular}
\caption{The list of domain words ultimately removed from the cleaned dataset.}
\end{figure}
\\
By the end of the process, we have just over 270,000 cleaned sparse vectors prepped for clustering. The nature of the dataset is at this point primarily descriptor words, which is essential to the clustering algorithm's efficacy.
\begin{figure}[h!]
\centering
\includegraphics[width=6.3cm]{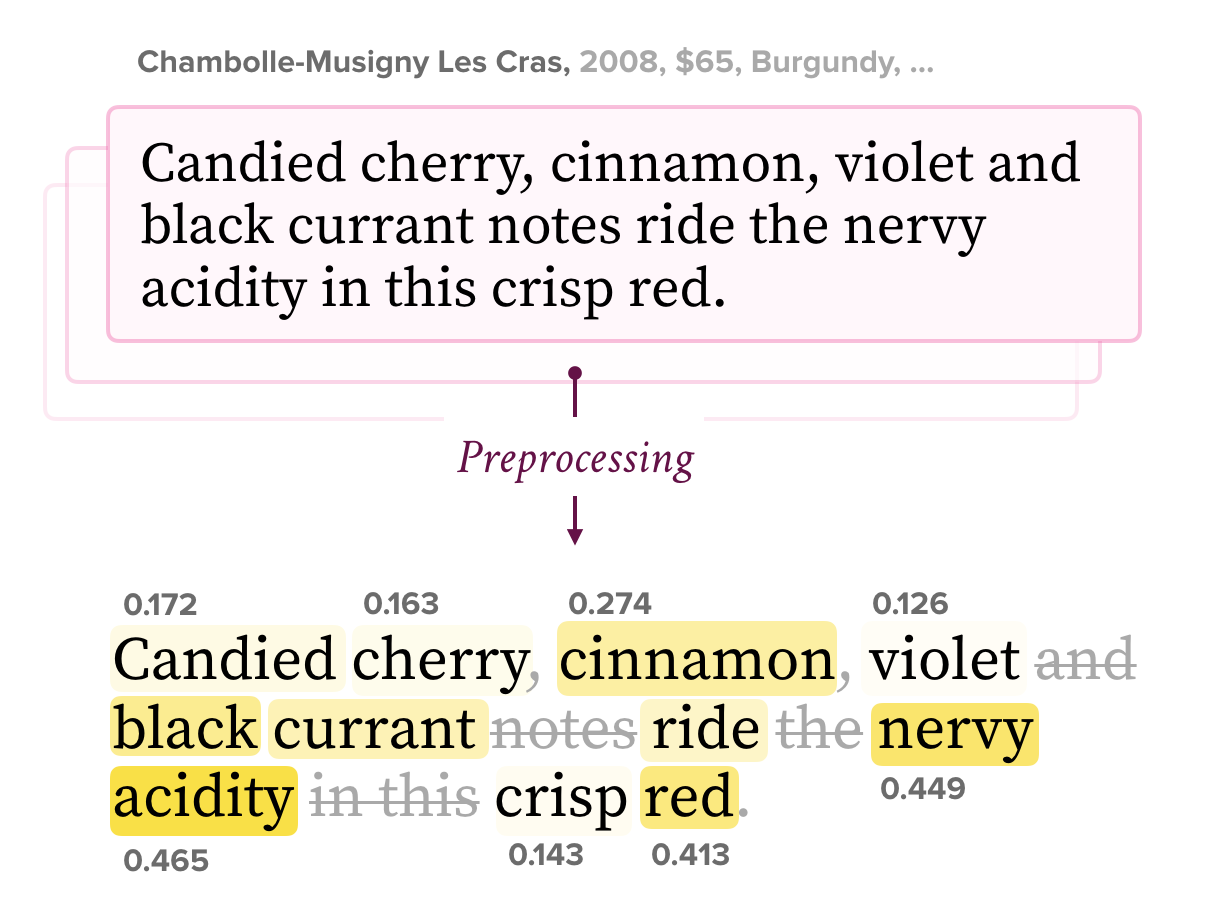}
\caption{A review for a wine (top) and it's corresponding cleanup (bottom), with some TF-IDF values labeled. The saturation of the color corresponds to the relative "strength" of the TF-IDF value for each word.}
\end{figure}

\section{Methodology}

Our methodology was divided into two sections: First we developed a clustering algorithm to group together wines based on similar wine reviews. This required the use of feature extraction tools. Secondly, we designed and implemented an optimization function that within a group of recommended wines returns to the user a wine with a maximized price-quality ratio. We will now more closely explore each of these two sections: 

\subsection{Clustering} 

After scraping our data from WineSpectator.com, we filtered out the reviews for their the descriptive words. As described in the previous section, we concentrated closely on keeping the adjectives and descriptive phrases of reviews in the description. After our filtering step, we were left with the ‘key’ descriptive features of each review. Using these isolated words, we then created a frequency matrix for the corpus of the reviews, normalizing each frequency array using TF-IDF. This process is visualized for a particular review in Fig. 3. We also experimented with GLoVE word embeddings over TF-IFD, which is discussed in detail in Section V.  %On this corpus, we then converted each array corresponding to a review’s word frequency to a glove-vector. Note that glove vectors are a method through which to measure the linguistic and semantic similarity between groups of words in a document. The program for converting our features to glove vectors was derived directly from Pennington, Socher and Manning’s paper [11]. 

Using each example’s sparse vectors as coordinates, we experimented with two clustering algorithms; K-means and EM. Clustering is imperative to our endeavor because it enables us to reduce our runtimes dramatically during the recommendation step. By limiting search for optimal wine… 

 Due to long runtimes, we used our k-means implementation to derive the optimal cluster count $k$, using the Elbow method as described by Kodinariya and Makwana [12]. We found that after $k=32$ increases in optimality were insignificant when compared to time and computational cost. 
 
\begin{figure}[h!]
\centering
\includegraphics[width=2.5in]{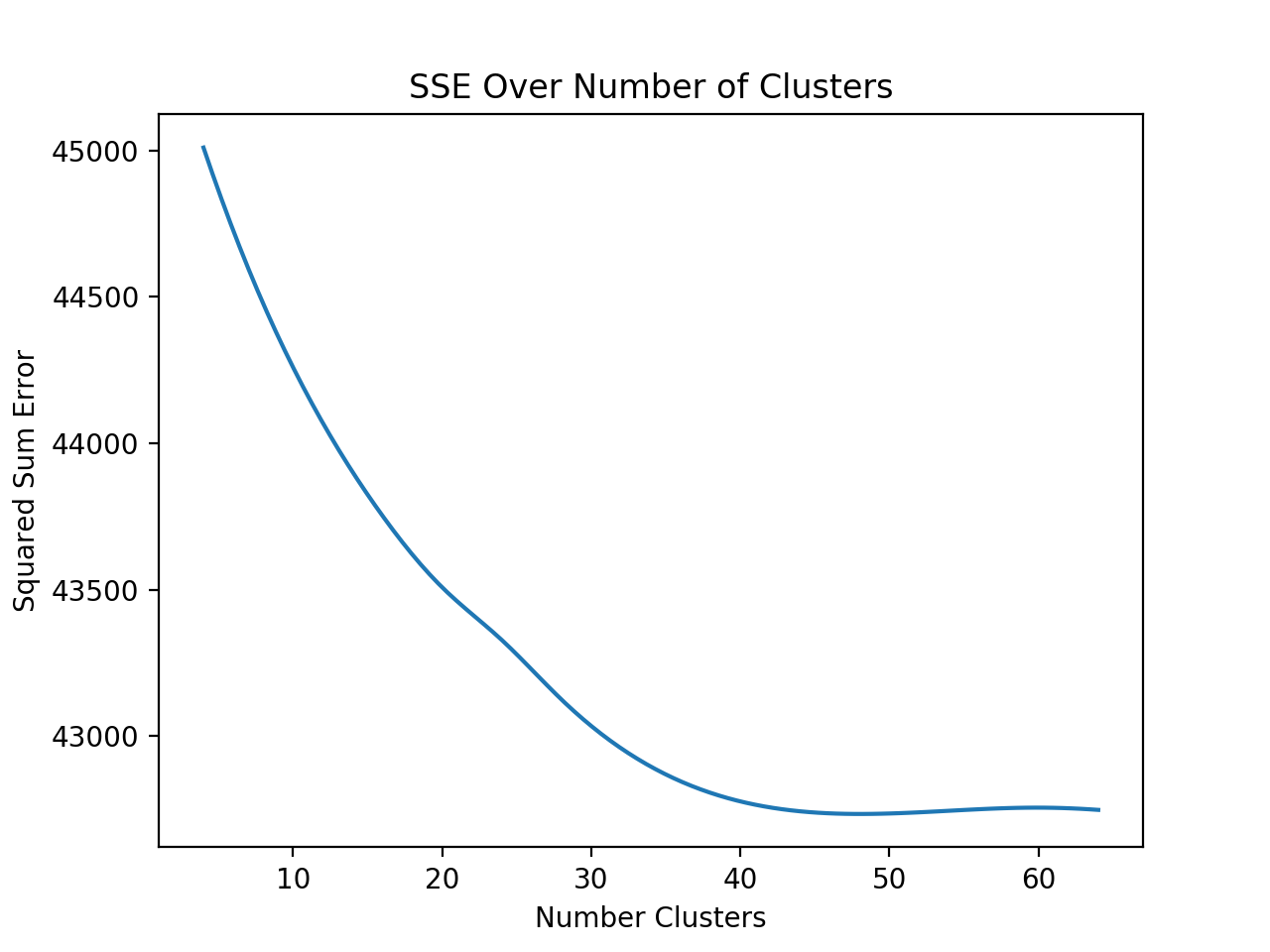}
\caption{SSE over Number of Clusters}
\end{figure}

Finally, we developed an algorithm for recommendation using these clusters. The algorithm looks at the history of a user's wine consumption. Using those wines, we sample from a Multinomial distribution where the probability of each of the k outcomes are dependent on the wines in a cluster the user liked and didn't like. If $H$ is the set of wines a user has tried, $x_k$ is the percentage of the user’s positively reviewed wines in cluster $k$, $y_k$ is the raw count of wines in the user’s history in cluster $k$, and $z_k$ is the percentage of the user’s negatively reviewed wines in cluster $k$, the probability of selecting some cluster $k$ is as follows:

$$    p_k= \frac{x_ky_k(1-z_k)}{\sum_{i=1}^{|H|} x_iy_i(1-z_i)} 
$$

The sample is then used to select the cluster in which we will search for a recommendation. From this cluster, a wine from the user's history is sampled at random with added multivariate Gaussian noise. This sample then serves as our benchmark coordinate. In order to take advantage of EM's soft clustering property, we then check the benchmark coordinate’s likelihood of being in each of the $k$ clusters we have defined via the multivariate Gaussian probability density function. If the two highest probabilities are close to equivalent, then we expand our search space to include both of the aforementioned clusters. 

\subsection{Selection Optimization}

Once the search spaces are defined, we iterate through each of the examples located in the target clusters and return to the user that which minimizes the following cost function, where $\lambda$ is some tunable hyperparameter scaling the weight of wine similarity:

$$ J(\pmb w, \pmb w') = \frac{\text{quality}(\pmb  w')}{\text{price}(\pmb  w')} + \lambda ||w - w'||_2
$$

Here, $\pmb w$ is the sampled history wine, and $\pmb w'$ is each of the candidate wines up for selection. $Quality$ is defined as the score of the wine as given by the sommeliers at WineSpectator.com, and $price$ is defined as the market price of the wine. The similarity function is simply Euclidian distance. We initially ran tests using cosine similarity, but ultimately settled on Euclidian distance because it is the method used by the clustering algorithms to evaluate closeness of data points.

In order to prevent the scenario in which a user becomes trapped in a single cluster, this current iteration currently returns 3 “Bets” and 1 “Wildcard.” The algorithm for the selection of a Wildcard is the same, except the multivariate Gaussian noise is added with a much broader covariance matrix.

\subsection{Other Considerations} 

For the purposes of training our model, we were required to create a proxy for a user’s history. Using terminology from the paper published by Netflix, we will refer to this as a “cold start.” For the demonstration and user tests, we drafted a variant of the application capable of short circuiting the prediction algorithm. We tried two approaches: artificial history generation and representative coordinate sampling. In both methods, we had the user fill out a questionnaire detailing their “ideal” wine. The artificial history generation implementation then looked at the top five wines with the highest TF-IDF values for each of the words and placed them in the history as if the user had given positive feedback to each of these wines. This was ineffective due to the fact that the wines were not necessarily representative of the clusters in which they were assigned. Thus we opted for representative coordinate sampling.

In the second approach, we took the response from the user and looked not at the wines, but the cluster centroids themselves. Because a centroid is representative of the wines in its cluster, we wrote a script capable of compiling a record of the top 10 TF-IDF valued indices in each cluster. We then iterated through the response of the user and matched their selected keywords to their respective clusters. The clusters that accumulated the most keywords were selected as target clusters. The benchmark coordinate is then not sampled from the user’s history, but sampled randomly from the centroids of the target clusters. By using the centroids of the clustering algorithm, we created a cold start algorithm capable of selecting the most representative wines for a user based on his or her responses to the questionnaire.

\section{Results \& Discussion} 
Our results will be divided into two sections: an evaluation of our clustering model based on exploratory analysis of our data, and the results of experiments we have run to evaluate the result of our prediction algorithm. These two metrics measure respectively how well our clustering algorithm works, and how effectively our optimization function is tuned to maximizing the likelihood that the user will purchase the recommended wine. 

\begin{figure}[h!]
\centering
\includegraphics[width=2.4in]{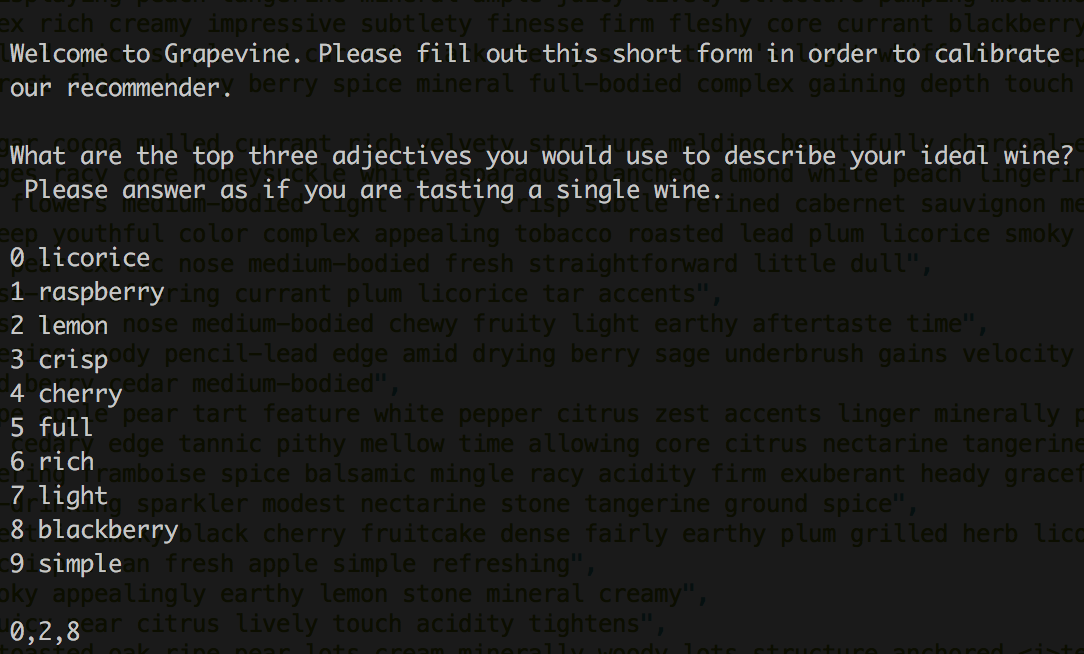}
\caption{Interface of the "Cold Start" Questionnaire.} 
\end{figure}
\begin{figure}[h!]
\center
\includegraphics[width=2.4in]{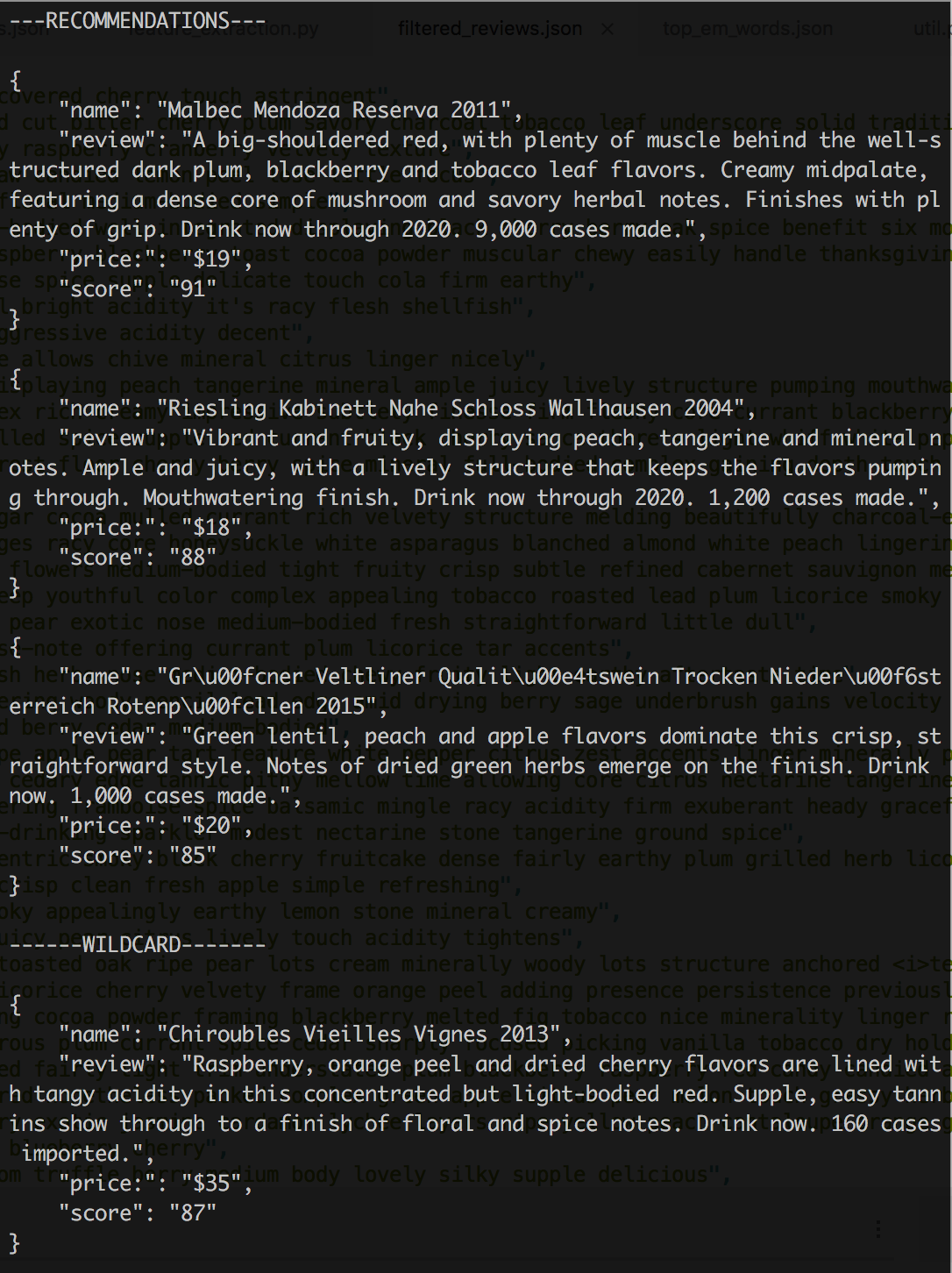}
\caption{Recommendations, given "Licorice," "Lemon," and "Blackberry."}
\end{figure}

	Exploratory Analysis of the Clusterings: We ran our model for 100 iterations, on a random selection of input preferences, keeping track on each iteration of both the recommended wine and the descriptive features of the selected wine. We then assigned a group of colleagues and peers to scan over the input and output of each iteration, and report whether the descriptive features of the output roughly match the randomly selected preferences. We chose to not do this evaluation ourselves for fear of researcher bias. We chose the following phrasing when prompting our colleagues (5 individuals) for their input: 

	\textit{Do the following descriptive words [referring to the output features of the recommended wine] resemble closely the former set of descriptive words [referring to the input preferences that our algorithm was initialized with]? }

	The result of our analysis showed that of the 100 iterations, 91\% of times the output wine descriptions were congruent with the input preferences. 

	%As part of our exploratory data analysis, we also sought to compare how the use of glove vectors improved the overall accuracy of the clustered models. In order to quantify this improvement, we ran our model for 50 iterations, extracting features from the wine reviews simply by using TF-IDF (without the use of the glove algorithm). We then again compared the output of the features predicted by the model with the input preferences that we initialized the model with. Asking our colleagues (3 persons, different than the original set of individual we asked) to compare the input and output data as before. We now observed that the output wine descriptions were only congruent with the input preferences 84\% of the time. Naturally, the difference in accuracy could result from the narrow set of individual we asked to help us determine the accuracy of our model. Nonetheless, this rough approximation lends some preliminary credence to the use of glove vectors as a tool to improve our clustering method. 
	
	\begin{figure}[h!]
	\center 
	\includegraphics[width=2in]{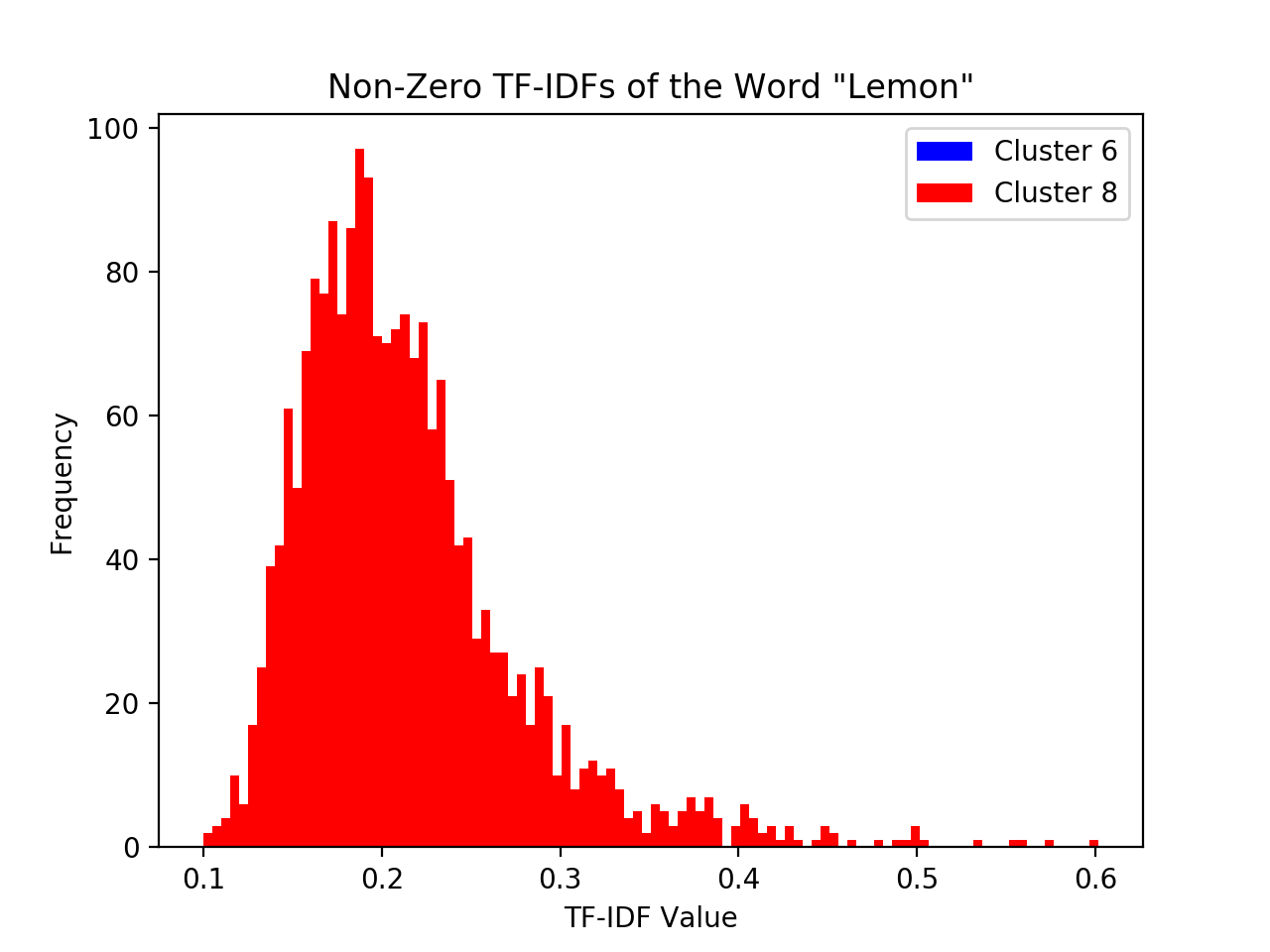}
	\caption{Frequency of Non-Zero TF-IDF values for the Word "Lemon."} 
	\end{figure}
	\begin{figure}[h!]
	\center
	\includegraphics[width=2in]{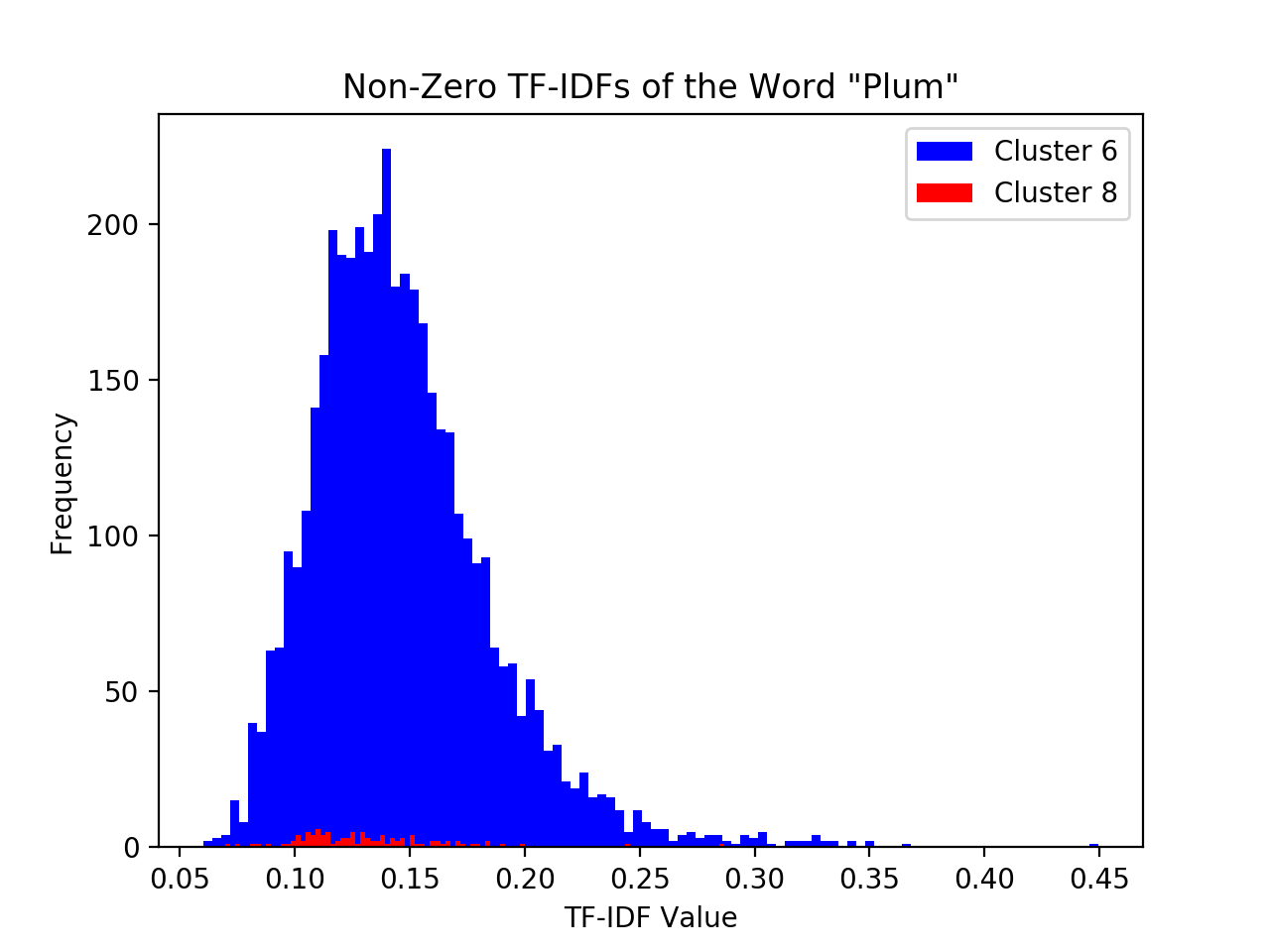}
	\caption{Frequency of Non-Zero TF-IDF values for the Word "Plum."}
	\end{figure}
	
	Generally, we observed that our clusters were able to isolate individual words very well. The following graphs demonstrate how clusters represent a clear overwhelming proportion of certain words (in this case comparing the word lemon and plum), indicating that the clustering algorithm works as expected.

	Experiment: After our exploratory analysis, we conducted a series of experiments to determine whether our optimization function was correctly tuned to maximize the likelihood that an individual would purchase the recommended wine our algorithm suggests. Given the lack of labeled data, we again were required to conduct independent questionnaires to determine how well our algorithm was attuned to predicting optimal wines. 
	To gather data, we asked 25 friends to each run our model five times. Naturally, our experimental design is heavily flawed: for one we were not able to incorporate a control group, and the subjects were all (most likely) biased to provide us with data points that supported our algorithm. After the run of each iteration, we asked all the participants in our study group the following question:

	\textit{If you were to purchase wine, would you buy the recommended wine over the wine that you would normally purchase?} 

In total, we were able to gather 117 responses (some subjects did not run the algorithm for the total number of iterations we instructed them to). Out of these 117 responses, 65\% of the time subjects reported they would rather purchase the recommended wine than their regular choice. When asked why they would not purchase our recommendation more often, nearly all subjects  (92\%) responded that the recommended wine was too expensive. 
\\\\
Finally, we experimented with GLoVE word embeddings [13]. The GLoVE algorithm obtains vector representation for words in a corpus such that the dot product of any two word vectors tries to equal their probability of co-occurrence over the corpus. Thus, GLoVE vectors could capture relationships between different words used in the reviews, as opposed to just the relative importance of particular words, and therefore led to significantly more nuanced clustering.
\\\\
We trained GLoVE vectors of size 50 over our corpus of filtered reviews. We chose to train GLoVE vectors on our own corpus (as opposed to using pre-trained vectors over the English language, from corpora like Wikipedia and Twitter) because the intended meaning of the language in wine reviews is highly contextual and idiomatic (which is exactly what makes them inaccessible to the average person in the first place), and thus GLoVE vectors trained specifically within this space likely capture the intended meaning more precisely. Each review was then represented as a size $50n$ vector, where $n$ is the size of the vocabulary (in our case, 13,324). The resulting dimensionality of the design matrix made clustering efficiently incredibly challenging, and we resorted to performing mini-batch k-means with k = 12. \\\\ Still, the early results were promising: the most representative reviews of a number of clusters (i.e. those reviews closest to the centroid) contained several different adjectives with very similar meaning. In particular, one cluster captured $smokey$, $tobacco$, and $cigar$, another captured $woodsy$, $earthy$, and $mineral$, while another captured $soft$, $light$, and $delicate$.

\section{Conclusion \& Future Work}
Given the outline of the problems listed in the previous section, it is clear that more work remains to be done in terms of tuning the hyper-parameters of our model. The discrepancy we observed between the effective clustering of our wines, and the moderate performance of our overall prediction algorithm can be explained by the lack of result data. This makes it difficult for our model to learn the optimal trade-off between wine similarity and price-to-quality ratio. Future work will therefore be concentrated on gathering more user feedback data on the accuracy of our model predictions. One possible method of doing so is to run our model as part of a survey on Mechanical Turk. The survey would ask Mechanical Turk workers if they would be more likely to purchase the recommended wine over their normal wine selection. Naturally, one limitation of this approach is that Mechanical Turkers are not representative of the overall population, and perhaps not of the clientele who would be most likely to use this algorithm, potentially biasing our results. Aside from this, the algorithm has yielded promising results thus far and we look forward to future iterations on the subject matter.

\ifCLASSOPTIONcaptionsoff
  \newpage
\fi



%\bibliographystyle{IEEEtran}

\section{Contribution}

All of us equally contributed the project outline and algorithm design. Below are some things that individual group member's took a lead on: 

Geoffrey Angus: Augmented the winery scraper to handle custom input and connect directly to the review scraping pipeline. Wrote the entire review scraper pipeline from there, building in failure-redundancy and the ability to take in custom input. Scraped and compiled the data over the course of several weeks and then aggregated it into a review JSON file. Implemented much of the infrastructure required to run the demo version of the software through both artificial history generation and representative point sampling. Pair programmed the predictor algorithm. Brought together the components to make it work as a cohesive system. Drafted the Dataset and Features section of the final paper along with most of the figures in the final report.

Richard Diehl Martinez: Implemented the EM and K-means clustering algorithms and the pipeline that feeds in the data to the clustering functions. Also developed the outline for the feature extraction code. Designed the general layout for the Classes and Methods used in the data pipeline from the clustering to the prediction algorithm. Helped with developing and implementing the optimization function that finds the optimal wine within a cluster. Wrote a majority of the final paper.  

Roozbeh Mahdavian: Trained GLoVE vectors over the corpus, and re-architected the pipeline to support representing them in memory (by iteratively building and repacking sparse matrices) and training them via mini-batch Kmeans. Also developed the baseline scraper code, and designed the layout and all visualizations for the poster. Contributed to developing the optimization function and the clustering approach. Drafted the introduction and the GLoVE vector overview of the final paper.

%\begin{biography}[{\includegraphics[width=1in,height=1.25in,clip,keepaspectratio]{mshell}}]{Michael Shell}

\begin{IEEEbiography}[{\includegraphics[width=1in,height=1.25in,clip,keepaspectratio]{picture}}]{John Doe}
\blindtext
\end{IEEEbiography}


%\vfill


\end{document}